\pdfoutput=1
\documentclass[aps,nofootinbib,prd,showpacs,class-pre]{revtex4}
\usepackage{amsfonts}
\usepackage{amsmath}
\usepackage{amssymb}
\usepackage[english]{babel}
\usepackage{slashed}
\usepackage{graphicx, epsfig}
\usepackage{xcolor}
\usepackage{natbib}
\usepackage{relsize}
\usepackage{subfig}
\usepackage{appendix}

\DeclareMathOperator{\arccosh}{arccosh}

\newcommand{\eq}{\begin{equation}}
\newcommand{\be}{\begin{equation}}
\newcommand{\feq}{\end{equation}}
\newcommand{\ee}{\end{equation}}

\begin{document}

\begin{flushleft}
KCL-PH-TH/2015-40
\end{flushleft}

\title{Semiclassical solutions of generalized Wheeler-DeWitt cosmology}

\author{Marco de Cesare$^1$\footnote{marco.de$_{-}$cesare@kcl.ac.uk}, 
Maria Vittoria Gargiulo$^2$\footnote{mariavittoria.gargiulo@gmail.com},
Mairi Sakellariadou$^1$\footnote{mairi.sakellariadou@kcl.ac.uk}}

\affiliation{$^1$Department of Physics, King's College London,
  University of London, Strand, London WC2R 2LS, U.K.}
\affiliation{$^2$Dipartimento di Fisica, I.N.F.N., Universit\`a di
  Salerno, I-84100 Salerno, Italy}

\begin{abstract}
We consider an extension of WDW minisuperpace cosmology with
additional interaction terms that preserve the linear structure of the
theory. General perturbative methods are developed and applied to
known semiclassical solutions for a closed Universe filled with a
massless scalar. The exact Feynman propagator of the free theory is
derived by means of a conformal transformation in minisuperspace. As
an example, a stochastic interaction term is considered and first
order perturbative corrections are computed. It is argued that such an
interaction can be used to describe the interaction of the
cosmological background with the microscopic d.o.f. of the
gravitational field. A Helmoltz-like equation is considered for the
case of interactions that do not depend on the internal time and the
corresponding Green's kernel is obtained exactly.The possibility of
linking this approach to fundamental theories of Quantum Gravity is
investigated.
\end{abstract}


\maketitle

\section{Introduction}\label{sec:intro}

The Wheeler-DeWitt (WDW) equation represents the starting point of the
canonical quantization program, also known as geometrodynamics. It was
originally derived by applying Dirac's quantization scheme to
Einstein's theory of gravity, formulated in Arnowitt-Deser-Misner
(ADM) variables. It has the peculiar form of a timeless
Schr{\"o}dinger equation, whose solutions are functions defined on the
space of three-dimensional geometries
\be
H\psi=0~.
\label{wdw}
\ee
The equation is a constraint imposed on physical states, encoding the
invariance of the theory under reparametrization of the time
variable. Together with the (spatial) diffeomorphism constraint, it
expresses the principle of general covariance at the quantum
level. This approach has however several limitations that prevented
it from being accepted as a fully satisfactory theory of Quantum
Gravity in its original formulation\footnote{For a general account of the WDW theory and the
  problems it encounters see \emph{e.g.}
  Ref.~\cite{Rovelli:2015}.}. Nonetheless, some of the issues it raises, as for
instance the problem of time (\emph{i.e.} the lack of a universal
parameter used to describe the evolution of the gravitational field),
are actually problems encountered by all fundamental
(nonperturbative) approaches to quantum gravity.  Others stem instead
from the dubious mathematical structure of the theory. Among the most
essential ones in the latter category we mention the factor ordering
problem in the Hamiltonian constraint and the construction of a
physical Hilbert space.

Despite its limitations, WDW was proved to be a valuable instrument in
cases where the number of degrees of freedom is restricted \emph{a
  priori}, \emph{i.e.} in minisuperspace and midisuperspace models. In
particular, its application to cosmological settings has been
extremely useful to gain insight into some of the deep questions
raised by a quantum theory of the gravitational field, such as the
problem of time~\cite{Halliwell} and the occurrence of stable
macroscopic branches for the state of the Universe (as in the
consistent histories approach~\cite{Craig-Singh:2010}).

One of the fundamental aspects of WDW is its linearity. This is indeed
a consequence of the Dirac quantization procedure and a property of
any first-quantized theory. One can nonetheless find in the literature
several proposals for non-linear extensions of the geometrodynamics
equation. They are in general motivated by an interpretation of $\psi$
in Eq.~(\ref{wdw}) as a field operator (rather than a function)
defined on the space of geometries. This idea has been applied to
Cosmology and found concrete realization in the old baby Universes
approach~\cite{Giddings:1988}. More recently, a model motivated by Group
Field Theory (GFT) (see Ref.~\cite{Oriti:2014} for a recent outline) has been
proposed in Ref.~\cite{Calcagni:2012} that retains the same spirit, taking
Loop Quantum Cosmology (LQC)~\cite{AshtekarBojowald2003} as a starting
point\footnote{Yet another interpretation of the dynamical equation on
  minisuperspace has been recently advocated in Ref.~\cite{Oriti:2010}. According to this point of view, it should rather be
  interpreted as some analogue of a (non-linear) hydrodynamical
  equation, for which the superposition principle does not hold in
  general.}.

It is worthwhile stressing that the WDW theory can be recovered in the
continuum limit (see Ref.~\cite{Ashtekar:2006} for details) from a more
fundamental theory such as LQC. In this sense it can be interpreted as
an effective theory of Quantum Cosmology, valid at scales such that the
fundamentally discrete structure of spacetime cannot be
probed. Therefore WDW, far from being of mere historical
relevance, is rather to be considered as an important tool to
extract predictions from cosmological models when a continuum, semiclassical behaviour is to be expected.

In what follows we consider as a starting point the Hamiltonian
constraint of LQC, leading to the free evolution of the Universe, as
discussed in the GFT-inspired model ~\cite{Calcagni:2012}.
Non-linear terms are allowed, which can be interpreted as interactions between
disconnected homogeneous components of an isotropic Universe
(scattering processes describe topology change). Instead of resorting
to a third quantized formalism built on minisuperspace models, we
follow a rather conservative approach which does not postulate the
existence of Universes disconnected from ours. We therefore only take
into account linear modifications of the theory, so as to guarantee
that the superposition principle remains satisfied. Hence these extra
terms can be interpreted as self-interactions of the Universe, as well
as violations of the Hamiltonian constraint.

We aim at studying the dynamics in a regime such that the free LQC
dynamics (given by finite difference equation) can be approximately
described in terms of differential equations. In this regime, the
dynamics is given by a modified WDW equation. The additional
interactions should be such that deviations from the Friedmann
equation are small, and will therefore be treated as small
perturbations.

We consider a closed Friedmann-Lema\^{i}tre-Robertson-Walker (FLRW)
 Universe, for which
semiclassical solutions are known explicitly~\cite{Kiefer:1988}. Corrections to the solutions
arising from the extra linear interactions are obtained
perturbatively. Of particular interest is the case in which the
perturbation is represented by white noise. In fact, this could be a
way to model the effect of the discrete structure underlying spacetime
on the evolution of the macroscopically relevant degrees of
freedom. It should therefore be possible to make contact, at least
qualitatively, between the usual quantum cosmology and the GFT
cosmology, according to which the dynamics of the Universe is that of
a condensate of elementary spacetime constituents. Considering a white
noise term amounts to treating the vacuum fluctuations of the
gravitational field as a stochastic process, an approach motivated by
an analogy with Stochastic Electrodynamics (SED)~\cite{Lamb-shift-SED, de1983stochastic, Lamb-shift-spontaneous-emission-SED, QM-SED}. This will be done is
Section~\ref{section noise}. The analogy with SED has of course its limitations,
since in the case at hand there is no knowledge about the dynamics of
the vacuum, which should instead come from a full theory of Quantum
Gravity. In spite of this limitation, the methods developed here are
fully general, so as to allow for a perturbative analysis of the
solutions of the linearly modified WDW equation for any possible form
of the additional interactions.

In the present work we restrict to a closed Universe, for which wave
packet solutions were constructed in Ref.~\cite{Kiefer:1988}.
In the same work, it was shown that it is
possible to construct a quantum state whose evolution mimics a solution of the
classical Friedmann equation, and is given by a quantum superposition
of two Gaussian wave packets (one for each of the two phases of the
Universe, expanding and contracting) centered on the classical
trajectory.  We build on the results of Ref.~\cite{Kiefer:1988},
considering the new interaction term as a perturbation, and thus
determining the corrections to the motion of the wave packets.

The rest of the paper is organized as follows: In Sec.~II, we show how
our model is motivated by LQC and GFT inspired
cosmology~\cite{Calcagni:2012} in the continuum limit (large
volumes).  In Sec.~III, we review the construction of the wave packet
solutions done in Ref.~\cite{Kiefer:1988}. These solutions will be used in
the subsequent sections as the unperturbed states describing the
evolution of a semiclassical Universe.  In Sec.~IV, using the scalar
field to define an internal time, we develop a framework for
time-independent perturbation theory. The inverse of the Helmoltz
operator corresponding to the free theory is computed exactly and
turns out to depend on a real parameter, linked to the choice of the
boundary conditions at the singularity. An analogue of Ehrenfest's
theorem for the evolution equation of the expectation values of
observables is then given.  In Sec.~V, the Feynman propagator of the
WDW free field operator is evaluated exactly. This is accomplished by
using a conformal map in minisuperspace, which reduces the problem to
that of finding the Klein-Gordon propagator in a planar region with a
boundary.  In Sec.~VI, we consider the case in which the additional
interaction is given by white noise.  In Sec.~VII, we discuss the
r\^ole of the Noether charge and the choice of the inner product.
Finally, we review our results and their physical implications in
Sec.~VIII.


\section{From the LQC free theory to WDW}\label{section 2}

Let us briefly show how the WDW equation is recovered from the
Hamiltonian constraint of LQC following
Ref.~\cite{Ashtekar:2006}~\footnote{The actual way in which WDW
  represents a large volume limit of LQC is put in clear mathematical
  terms in Refs.~\cite{Ashtekar:2007, Corichi:2007}, where the
  analysis is based on a special class of solvable models (sLQC).}.
For our purposes it is convenient to consider a massless scalar field
$\phi$, minimally coupled to the gravitational field. This choice has
the advantage of allowing for a straightforward deparametrization of
the theory, thus defining a clock. The Hamiltonian constraint has the
general structure~\cite{Ashtekar:2006, Calcagni:2012}
\be
\hat{\mathcal{K}}\psi(\nu,\phi)\equiv-B(\nu)
\left(\Theta+\partial_{\phi}^2\right)\psi(\nu,\phi)=0~,
\ee
where $\psi$ is a wave function on configuration space, $\Theta$ is a
finite difference operator acting on the gravitational sector in the
kinetic Hilbert space of the theory $\mathcal{H}^{\rm g}_{\rm kin}$,
and $\nu$ denotes the generic eigenvalue of the volume operator. The
discreteness introduced by the LQC formulation does not affect the
matter sector, which is still the same as in the continuum WDW quantum
theory. The gravitational sector of the Hamiltonian constraint
operator of LQC in ``improved dynamics''~\footnote{In the framework of
  LQC, the Hamiltonian constraint contains the gravitational
  connection $c$. However, only the holonomies of the connections are
  well defined operators, hence to quantize the theory we replace $c$
  by $\sin\bar\mu c/\bar \mu$, where $\bar\mu$ represents the
  ``length'' of the line segment along which the holonomy is
  evaluated.  Originally $\bar\mu$ was set to a constant $\mu_0$,
  related to the area-gap. To cure severe issues in the ultraviolet
  and infrared regimes which plague the $\mu_0$ quantization, a new
  scheme called ``improved dynamics'' was
  proposed~\cite{Ashtekar:2006improved}. In the latter, the dimensionless
  length of the smallest plaquette is $\bar\mu$. }, reads 
\be
-B(\nu)\Theta\,\psi(\nu,\phi)\equiv
A(\nu)\psi(\nu+\nu_0,\phi)+C(\nu)\psi(\nu,\phi)+D(\nu)\psi(\nu-\nu_0,\phi)~,
\ee 
where the finite increment $\nu_0$ represents an elementary volume unit and $A, B, C, D$ are
functions which depend on the chosen quantization scheme.  In order to
guarantee that $\Theta$ is symmetric\footnote{In the large volume
  limit it will be formally self-adjoint w.r.t. the measure
  $B(\nu)\mbox{d}\nu$.} in $\nu$, the coefficients must satisfy the
$D(\nu)=A(\nu-\nu_0)$ condition~\cite{Calcagni:2012}; it holds in both
the $k=0$ and the $k=1$ case.

It is a general result of LQC that WDW can be recovered in the
continuum (\emph{i.e.} large volume) limit~\cite{Ashtekar:2007}.
 In particular, it was shown in Ref.~\cite{Ashtekar:2006} that for
 $k=1$ one recovers the Hamiltonian constraint of
 Ref.~\cite{Kiefer:1988}. In fact it turns out that $\Theta$ can be
 expressed as the sum of the operator ($\Theta_0$) relative to the
 $k=0$ case and a $\phi$-independent potential term (\emph{i.e.}
 diagonal in the $\nu$ basis) as
\be\label{kinetic operator}
\Theta=\Theta_0+\frac{\pi G l_0^2\gamma^2}{3 K^{4/3}}|\nu|^{4/3}~.
\ee
In the above expression $K=\frac{2\sqrt{2}}{3\sqrt{3\sqrt{3}}}$,
$\gamma$ is the Barbero-Immirzi parameter of Loop Quantum Gravity
(LQG), $G$ the gravitational constant and $l_0$ is the cube root of
the fiducial volume of the fiducial cell on the spatial manifold in
the $k=1$ model. The latter can be formally sent to zero in order to
recover the $k=0$ case.

Restricting to wave functions $\psi(\nu)$ which are smooth and slowly
varying in $\nu$, we obtain the WDW limit of the Hamiltonian
constraint
\be
\Theta_0\psi(\nu,\phi)=-12\pi G (\nu\partial_{\nu})^2\psi(\nu,\phi)~,
\ee
which is exactly the same constraint one obtains in WDW theory.  Thus,
LQC naturally recovers the factor ordering (also called
factor ordering, in the sense that the quantum
constraint operator is of the form $G^{AB}\nabla_A\nabla_B$, where
$G^{AB}$ is the inverse WDW metric and $\nabla_A$ denotes the covariant
derivative associated with $G_{AB}$) which was obtained in
Ref.~\cite{Halliwell} under the requirement of field reparametrization
invariance of the minisuperspace path-integral.  Since $\nu$ represents a
proper volume, it is proportional to the volume of a comoving cell
with linear dimension equal to the scale factor $a$
\be
\nu\propto a^3~.
\ee
Introducing the variable $\alpha=\log a$, we rewrite the constraint
operator $\hat{\mathcal{K}}$ as
\be \hat{\mathcal{K}}
=e^{-3\alpha}\left(\frac{\partial^2}{\partial\alpha^2}-\frac{\partial^2}{\partial\phi^2}-e^{4\alpha}\right)~.
\ee
Let us consider a modified dynamical equation of the form \be
\hat{\mathcal{K}}\psi(\nu,\phi)+\sum_{\nu^{\prime}}\int\mbox{d}\phi^{\prime}
\; g(\nu,\nu^{\prime};\phi,\phi^{\prime})
\psi(\nu^{\prime},\phi^{\prime})=0~, \ee
which, in the continuum limit $\nu\gg\nu_0$, leads to
\be\label{equation from GFT}
\left(\frac{\partial^2}{\partial\alpha^2}-\frac{\partial^2}{\partial\phi^2}-e^{4\alpha}\right)\psi=-e^{3\alpha}\int
\mbox{d}\alpha^{\prime}\,\mbox{d}\phi^{\prime}\;
\left.{\mbox{d}\nu\over\mbox{d}\alpha}\right|_{\alpha^{\prime}}g\psi~.
\ee
We will assume that the interaction $g$ is local in minisuperspace,
\emph{i.e.}
$g(\nu,\nu^{\prime};\phi,\phi^{\prime})=g(\nu,\phi)\delta(\nu-\nu{\prime})\delta(\phi-\phi')$. Given
the properties of the Dirac delta function
\be
\delta(\nu-\nu^{\prime})=\frac{\delta(\alpha-\alpha^{\prime})}{\left.{\mbox{d}\nu\over\mbox{d}\alpha}\right|_{\alpha^{\prime}}}~,
\ee
Equation~(\ref{equation from GFT}) reduces to a Klein-Gordon equation
with space and time dependent potential $e^{3\alpha} g(\alpha,\phi)$;
note that $g$ is kept completely general.

Since the interaction term represented by the r.h.s. of
Eq.(\ref{equation from GFT}) is unknown, we cannot determine an exact
solution of the equation without resorting to a case by case
analysis. However, since the solutions of the WDW equation in the
absence of a potential are known explicitly, we will adopt a
perturbative approach. The method we develop is fully general and can
thus be applied for any possible choice of the function $g$.

We formally expand the wave function and the WDW operator in terms of a dimensionless  parameter $\lambda$ (that serves book-keeping purposes and will be eventually set equal to $1$).
\begin{align}
\psi=\psi^{(0)}+\lambda\psi^{(1)}+\dots~,\\
\mbox{and}\ \ \hat{T}=\hat{T}_{0}+\lambda e^{3\alpha}\int g~,
\end{align}
with the definitions
\be\label{constraint operator}
\hat{T}_{0}\equiv\left(\frac{\partial^2}{\partial\alpha^2}-\frac{\partial^2}{\partial\phi^2}-e^{4\alpha}\right).
\ee
We therefore have
\begin{align}
\hat{T}_{0}\psi^{(0)}&=0~,\\
\hat{T}_{0}\psi^{(1)}&=-e^{3\alpha}g\,\psi^{(0)}\label{first order perturbation}~.
\end{align}
The zero-th order term $\psi^{(0)}$ is a solution of the wave equation with an exponential potential, which was obtained in Ref.~\citep{Kiefer:1988} and will be reviewed in the next section. If we were able to invert $\hat{T}_{0}$, we would get the wave function corrected to first order. However, finding the Green's function is not straightforward in this case as it would be for $k=0$ (where the kinetic operator is just the d'Alembertian, whose Green's kernels are well known for all possible choices of boundary conditions). Moreover, as for the d'Alembertian, the Green's kernel will depend on the boundary conditions. The problem of determining which set of boundary conditions is more appropriate depends on the physical situation we have in mind and will be dealt with in the next sections.

At this point, we would like to point out the relation between our approach and the model in Ref.~\cite{Calcagni:2012}, where a third quantization perspective is assumed. The action on minisuperspace that was considered in Ref.~\cite{Calcagni:2012} reads\footnote{The model was originally formulated for $k=0$, but it admits a straightforward generalization to include the case $k=1$.}
\be\label{action}
S[\psi]=S_{\rm free}[\psi]+\sum_{n}\frac{\lambda_{n}}{n!}\sum_{\nu_1\dots\nu_{n}}\int\mbox{d}\phi_1\dots\mbox{d}\phi_{n}\; f^{(n)}(\nu_i,\phi_i)\prod_{j}\psi(\nu_{j},\phi_{j})~,
\ee
where the first term gives the dynamics of the free theory, namely a homogeneous and isotropic gravitational background coupled to a massless scalar field 
\be
S_{\rm free}[\psi]=\sum_{\nu}\int\mbox{d}\phi\;\psi(\nu,\phi)\hat{\mathcal{K}}\psi(\nu,\phi)~;
\ee
$\hat{\mathcal{K}}$ is the Hamiltonian constraint in LQC and the terms containing the functions $f^{(n)}$ represent additional interactions that violate the constraint. This is a toy model for Group Field Cosmology~\citep{Calcagni:2012}, given by a GFT with group $G=U(1)$  and $\nu$ a Lie algebra element. The free dynamics depends on the specific LQC model adopted. However, the continuum limit should be the same regardless of the model considered and must give the WDW equation for  the corresponding three-space topology. The WDW approach to quantum cosmology should therefore be interpreted as an effective theory, valid at scales such that the discreteness introduced by the polymer quantization cannot be probed. We will show that, even from this more limited perspective, the action Eq.~(\ref{action}) leads to novel effective theory of quantum cosmology that represent modifications of WDW. 

It was hinted in Ref.~\cite{Calcagni:2012} that the additional interactions could also be interpreted as interactions occurring between homogeneous patches of an inhomogeneous Universe. Another possible interpretation is that they actually represent interactions among different, separate, Universes. The latter turns out to be a natural option in the framework of third quantization (see Ref.~\cite{Isham:1992} and references therein), which naturally allows for topology change\footnote{For an example of topology change in the old ``Baby Universes'' literature see \emph{e.g.} Ref.~\cite{Giddings:1988}.}. 

We restrict our attention to the quadratic term, that can be interpreted as a self interaction of the Universe. This is in fact worth considering even without resorting to a third-quantized cosmology and can in principle be generalized to include non-locality in minisuperspace (this situation will not be dealt with in the present work).


\section{Analysis of the unperturbed case}\label{Section unperturbed}

We review here the construction of wave packet solutions for the cosmological background presented in Ref.~\cite{Kiefer:1988}. The Wheeler-DeWitt equation for a homogeneous and isotropic Universe (compact spatial topology, $k=1$) with a massless scalar field is 
\be\label{WDW equation}
\left(\frac{\partial^2}{\partial\alpha^2}-\frac{\partial^2}{\partial\phi^2}-e^{4\alpha}\right)\psi(\alpha,\phi)=0~.
\ee
We impose the boundary condition
\be\label{boundary condition}
\lim_{\alpha\to\infty} \psi(\alpha,\phi)=0~,
\ee
necessary in order to reconstruct semiclassical states describing the dynamics of a closed Universe, since regions of minisuperspace corresponding to arbitrary large scale factors are not accessible.

Equation~(\ref{boundary condition}) can be solved by separation of variables
\be\label{sol-psi}
\psi_{k}(\alpha,\phi)=N_k\, C_{k}(\alpha)\varphi_{k}(\phi)~,
\ee
leading to
\begin{align}
\frac{\partial^2}{\partial\phi^2}\varphi_{k}+k^2\varphi_{k}&=0~,\label{EQUATION
  FOR
  PHI}\\ \frac{\partial^2}{\partial\alpha^2}C_{k}-(e^{4\alpha}-k^2) C_{k}&=0~.\label{EQUATION
  FOR ALPHA}
\end{align}
Note that $\psi_{k}(\alpha,\phi)$ stands for the solution of
Eq.~(\ref{WDW equation}) and should not be confused with the Fourier
transform of $\psi$, for which we will use instead the notation
$\tilde{\psi}(\alpha,k)$.  In Eq.~(\ref{sol-psi}) above, $N_k$ is a
normalization factor that depends on $k$, whose value will be fixed
later.

Let us proceed with the solutions of Eqs.~(\ref{EQUATION FOR PHI}),
(\ref{EQUATION FOR ALPHA}).  Equation~(\ref{EQUATION FOR PHI}) yields
complex exponentials as solutions
\be
\varphi_{k}=e^{ik\phi}~,
\ee
while Eq.~(\ref{EQUATION FOR ALPHA}) has the same form as the
stationary Schr\"odinger equation for a non-relativistic particle in
one dimension, with potential $V(\alpha)= e^{4\alpha}-k^2$ and zero
energy. One then easily remarks that the particle is free for
$\alpha\rightarrow -\infty$, whereas the potential barrier becomes
infinitely steep as $\alpha$ takes increasingly large positive
values. Given the imposed boundary condition, Eq.~(\ref{EQUATION FOR
  ALPHA}) admits as an exact solution the modified Bessel function of
the second kind (also known as MacDonald function) \be\label{MacDonald
  function} C_{k}(\alpha)=K_{ik/2}\left(\frac{e^{2\alpha}}{2}\right)~.
\ee
Wave packets are then constructed as linear superpositions of the
(appropriately normalized) solutions
\be\label{Wave Packet}
\psi(\alpha,\phi)=\int_{-\infty}^{\infty}\mbox{d}k\;\psi_{k}(\alpha,\phi)A(k)~,
\ee
with an appropriately chosen amplitude $A(k)$.
Since the $C_{k}(\alpha)$ are improper eigenfunctions of the
(one-parameter family of) Hamitonian operator(s) in Eq.~(\ref{EQUATION
  FOR ALPHA}), they do not belong to the space of square integrable
functions on the real line $L^{2}(\mathbb{R})$. However, it is still
possible to define some sort of normalization by fixing the
oscillation amplitude of the improper eigenfunctions for
$\alpha\to-\infty$. For this purpose, we recall the WKB expansion of
Eq.~(\ref{MacDonald function})
\be
K_{ik/2}\left(\frac{e^{2\alpha}}{2}\right)
=\frac{\sqrt{\pi}}{2}e^{-k\frac{\pi}{4}}(k^2-e^{4\alpha})^{-1/4}
\cos\left(\frac{k}{2}\arccosh\frac{k}{e^{2\alpha}}-\frac{1}{2}
\sqrt{k^2-e^{4\alpha}}
-\frac{\pi}{4}\right)~,
\ee
a very accurate approximation for large values of $k$. We therefore set
\be
\psi_{k}(\alpha,\phi)=e^{k\frac{\pi}{4}}\sqrt{k}\;K_{ik/2}
\left(\frac{e^{2\alpha}}{2}\right)e^{ik\phi}~,
\ee
which for large enough values of $k$ gives elementary waves with the
same amplitude to the left of the potential barrier. Note that for
small $k$, the amplitude $A(k)$ will still exhibit dependence on $k$.

\begin{figure}
\centering
\begin{minipage}{0.45\textwidth}
\centering
 \includegraphics[width=\columnwidth]
    {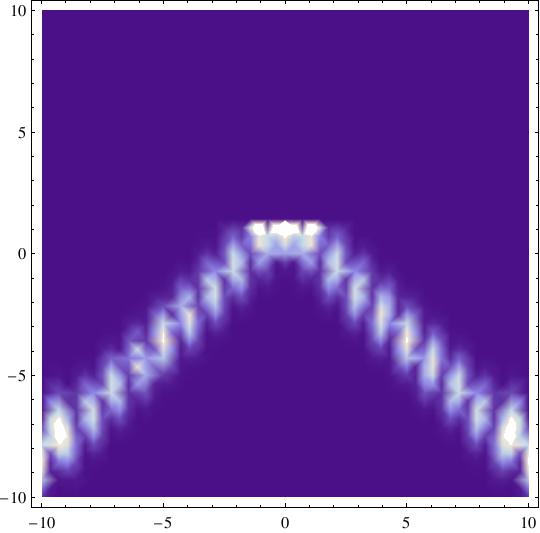}
    \caption{Absolute square of the ``wave function'' of the Universe
      corresponding to the choice of parameters $b=1$,
      $\overline{k}=10$. Lighter shades correspond to larger values of
      the wave function. During expansion and recollapse the evolution
      of the Universe can be seen as a freely propagating wave
      packet. From the plot it is also evident the reflection against
      the potential barrier at $\alpha_{\overline{k}}=\frac{1}{2}\log
      \overline{k}$, where the wave function exhibits a sharp peak.}\label{Semiclassical}
      \end{minipage}\hfill
\begin{minipage}{0.45\textwidth}
\centering
\includegraphics[width=\columnwidth]{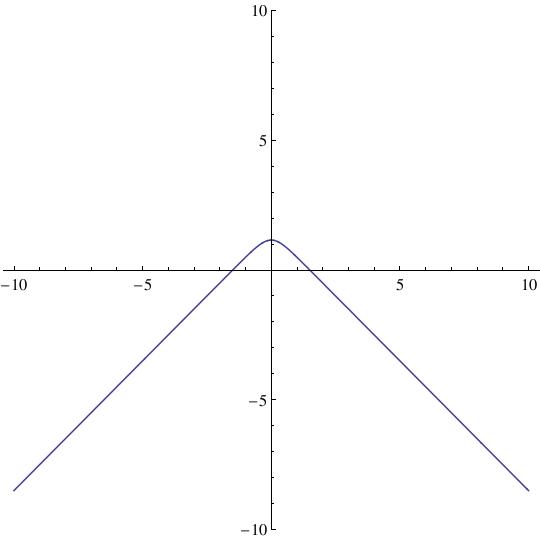}
    \caption{Classical trajectory of the Universe in
      minisuperspace. It is closely followed by the peak of
      semiclassical states, as we can see by comparison with
      Fig.~\ref{Semiclassical}.}
   \label{ClassicalTrajectory}
   \end{minipage}
\end{figure}

Let us assume a Gaussian profile for the amplitude $A(k)$ in
Eq.~(\ref{Wave Packet})
\be
A(k)=\frac{1}{\pi^{1/4}\sqrt{b}}e^{-\frac{(k-\overline{k})^2}{2b^2}}~,
\ee
where $\overline{k}$ should be taken large enough so as to guarantee
the normalization of the function $\psi_{k}(\alpha,\phi)$. One then
finds that the solution has the profile shown in the
Fig.~\ref{Semiclassical}.  The solution represents a wave packet that
starts propagating from a region where the potential vanishes
(\emph{i.e.} at the initial singularity) towards the potential barrier
located approximately at $\alpha_{\overline{k}}=\frac{1}{2}\log
\overline{k}$, from where it is reflected back. For such values of
$\overline{k}$ the wave packet is practically completely reflected
back from the barrier.  The parameter $b$ gives a measure of the
semiclassicality of the state, \emph{i.e.} it accounts for how much it
peaks on the classical trajectory.  We remark that the peak of the
wave packet follows closely the classical trajectory, 
\be\label{Classical Solution}
e^{2\alpha}= {\bar{k}\over \cosh (2\phi)}~.
\ee
and refer the
reader to Fig.~\ref{ClassicalTrajectory}.

\section{Time independent perturbation potential}

In the case in which the potential in the r.h.s of Eq.~(\ref{equation
  from GFT}) does not depend on the scalar field (that we interpret as
an \emph{internal time}) we can resort to time-independent
perturbation theory to calculate the corrections to the wave
function. Namely, we consider the representation of the operator $\hat
T_0$ defined in Eq.~(\ref{constraint operator}) in Fourier space, or
equivalently its action on a monochromatic wave, so that we are lead
to the Helmoltz equation in the presence of a potential
\be\label{homogeneous Helmoltz}
\left(\frac{\partial^2}{\partial\alpha^2}
+k^2-e^{4\alpha}\right)\tilde{\psi}^{(0)}(\alpha,k)=0~,
\ee
where the potential cannot be considered as a small perturbation with
respect to the standard Helmoltz equation. In fact, reflection from an
infinite potential barrier requires boundary conditions that are
incompatible with those adopted in the free particle case. We
therefore have to solve Eq.~(\ref{homogeneous Helmoltz}) exactly.

In order to compute the first perturbative corrections, we need to
solve Eq.~(\ref{first order perturbation}), which in Fourier space
leads to
\be\label{perturbation in Fourier space}
\left(\frac{\partial^2}{\partial\alpha^2}
+k^2-e^{4\alpha}\right)\tilde{\psi}^{(1)}(\alpha,k)
=U(\alpha)\tilde{\psi}^{(0)}(\alpha,k)~,
\ee
upon defining $U(\alpha)=-e^{3\alpha}g(\alpha)$.
Equation~(\ref{perturbation in Fourier space}) is easily solved once
the Green's function of the Helmoltz operator on the l.h.s is
known. The equation for the Green's function is
\be
\left(\frac{\partial^2}{\partial\alpha^2}
+k^2-e^{4\alpha}\right)G^{k}(\alpha,\alpha^{\prime})
=\delta(\alpha-\alpha^{\prime})~.
\ee
The homogeneous equation admits two linearly independent solutions,
which we can use to form two distinct linear combinations that satisfy
the boundary conditions at the two extrema of the interval of the real
axis we are considering. The remaining free parameters are then fixed
by requiring continuity of the function at $\alpha^{\prime}$ and the
condition on the discontinuity of the first derivative at the same
point.  

Equation~(\ref{homogeneous Helmoltz}) has two linearly independent
solutions $p(\alpha)$ and $h(\alpha)$ given by
\be
p(\alpha)=K_{i\frac{k}{2}}\left(\frac{e^{2\alpha}}{2}\right)
\ee
and
\be
h(\alpha)=I_{-i\frac{k}{2}}\left(\frac{e^{2\alpha}}{2}\right)
+I_{i\frac{k}{2}}\left(\frac{e^{2\alpha}}{2}\right)~.
\ee
We thus make the following ansatz
\be\label{ansatz}
G^{k}(\alpha,\alpha^{\prime})=\begin{cases} 
      \gamma\; p(\alpha) & \alpha\geq\alpha^{\prime} \\
      \delta\; h(\alpha)+\eta\; p(\alpha) & \alpha<\alpha^{\prime}~,
   \end{cases}
\ee
and note that $G^{k}(\alpha,\alpha^{\prime})$ satisfies the boundary
condition Eq.~($\ref{boundary condition}$) by construction.

Since we do not know what is the boundary condition for
$\alpha\to-\infty$, \emph{i.e.} near the classical
singularity\footnote{In fact, even popular choices like the
  Hartle-Hawking \emph{no-boundary} proposal~\cite{Hawking:1981,
    Hartle:1983} or the \emph{tunneling condition} proposed by
  Vilenkin~\cite{Vilenkin:1987}, only apply to massive scalar
  fields.}, we will not be able to fix the values of all constants
$\gamma,\delta,\eta$.
We will therefore end up with a one-parameter family of Green's functions.

The Green's function must be continuous at the point
$\alpha=\alpha^{\prime}$, hence
\be\label{eq1above} \gamma\; p(\alpha^{\prime})=\delta\;
h(\alpha^{\prime})+\eta\; p(\alpha^{\prime})~.  \ee
Moreover, in order for its second derivative to be a Dirac delta
functional, the following condition on the discontinuity of the first
derivative must be satisfied
\be\label{eq2above} \gamma\; p^{\prime}(\alpha^{\prime})-\delta\;
h^{\prime}(\alpha^{\prime})-\eta\; p^{\prime}(\alpha^{\prime})=1~.
\ee
Upon introducing a new constant $\Omega=\gamma-\eta$, we can rewrite
Eqs.~(\ref{eq1above}), (\ref{eq2above}) in the form of a Kramer's
system
\[
\begin{cases}
\Omega\; p(\alpha^{\prime})-\delta\; h(\alpha^{\prime})&=0~,\\
\Omega\; p^{\prime}(\alpha^{\prime})-\delta\; h^{\prime}(\alpha^{\prime})&=1~,
\end{cases}
\]
which admits a unique solution, given by
\begin{align}
\Omega&=-\frac{h(\alpha^{\prime})}{W(\alpha^{\prime})}~,\\
\delta&=-\frac{p(\alpha^{\prime})}{W(\alpha^{\prime})}~,
\end{align}
where
$W(\alpha^{\prime})=p(\alpha^{\prime})h^{\prime}(\alpha^{\prime})
-h(\alpha^{\prime})p^{\prime}(\alpha^{\prime})$
is the Wronskian. Hence the ansatz (\ref{ansatz}) is rewritten as
\[
G^{k,\eta}(\alpha,\alpha^{\prime})=\begin{cases}
p(\alpha)\left(\eta-\frac{h(\alpha^{\prime})}{W(\alpha^{\prime})}\right)
& \alpha\geq\alpha^{\prime}
\\ -\frac{p(\alpha^{\prime})}{W(\alpha^{\prime})}\; h(\alpha)+\eta\;
p(\alpha) & \alpha\leq\alpha^{\prime}~,
\end{cases}
\]
where we have explicitly introduced the parameter $\eta$ in our
notation for the Green's function in order to stress its
non-uniqueness. Note that the phase shift at $-\infty$ varies with
$\eta$.

Hence the solution of Eq.~(\ref{perturbation in Fourier space}) reads
\be\label{corrections tilde}
\tilde{\psi}^{(1)}(\alpha,k)=\int\mbox{d}\alpha^{\prime}\;
G^{k,\eta}(\alpha,\alpha^{\prime})U(\alpha^{\prime})
\tilde{\psi}^{(0)}(\alpha,k)~.
\ee
The possibility of studying the perturbative corrections using a
decomposition in monochromatic components is viable because of the
validity of the superposition principle. This is in fact preserved by
additional interactions of the type considered here, which violate the
constraint while preserving the linearity of the wave equation. Note
that in general, modifications of the scalar constraint in the WDW
theory would be non-linear and non-local, as for instance within the
proposal of Ref.~\cite{Oriti:2010}, where classical geometrodynamics
arises as the hydrodynamics limit of GFT. However, such non-linearities
would spoil the superposition principle, hence making the analysis of
the solutions much more involved.

The time dependence of the perturbative corrections is recovered by
means of the inverse Fourier transform of Eq.~(\ref{corrections
  tilde})
\be
\psi^{(1)}(\alpha,\phi)=\int\frac{\mbox{d}k}{2\pi}\; 
\tilde{\psi}^{(1)}(\alpha,k)e^{-ik\phi}~.
\ee

Expectation values of observables can be defined using the measure
determined by the time component of the Noether current $j_{\phi}$
(whose definition will be given in Sec.~\ref{Section on the charge}) as
\be \langle f\rangle=\int\mbox{d}\alpha\; f j_{\phi}~.  \ee 
Using the conservation of the Noether current we can then derive an
analogue of Ehrenfest theorem, namely
\be\label{Ehrenfest}
\frac{\mbox{d}}{\mbox{d}\phi}\langle f\rangle=\int\mbox{d}\alpha\;
\left( j_{\phi}\partial_{\phi} f +f
\partial_{\phi}j_{\phi}\right)=\int\mbox{d}\alpha\; \left(
j_{\phi}\partial_{\phi} f+f \partial_{\alpha}j_{\alpha}\right)~.  \ee
When the observable $f$ does not depend explicitly on the internal time
$\phi$, the first term in the integrand vanishes. Considering for
instance the scale factor $a=e^{\alpha}$ then, after integrating by
parts, we get 
\be \frac{\mbox{d}}{\mbox{d}\phi}\langle
a\rangle=-\int\mbox{d}\alpha\; e^{\alpha}j_{\alpha}~.  \ee 
In an analogous fashion, one can show that 
\be\label{Ehrenfest2}
\frac{\mbox{d}}{\mbox{d}\phi}\langle a^2\rangle=-2\int\mbox{d}\alpha\;
e^{2\alpha}j_{\alpha}~.  \ee 
Formulae (\ref{Ehrenfest}) and (\ref{Ehrenfest2}), besides their
simplicity, turn out quite handy for numerical computations,
especially when dealing with time independent observables. In fact
they can be used to compute time derivatives of the averaged
observables without the need for a high resolution on the $\phi$ axis,
\emph{i.e.} they can be calculated using data on a single
time-slice. Expectation values can therefore be propagated forwards or
backwards in time by solving first order ordinary differential
equations.

\section{First perturbative correction for time-dependent potentials}

In the previous section we considered a time-independent perturbation,
which can be dealt with using the Helmoltz equation. This is in
general not possible when the perturbation depends on the internal
time. In order to study the more general case we have to resort to
different techniques to find the exact Green's function of the
operator $\hat{T}_{0}$.  Since the potential $e^{4\alpha}$ breaks
translational symmetry, Fourier analysis, which makes the
determination of the propagator so straight-forward in the $k=0$ case
(where the Hamiltonian constraint leads to the wave equation), is of
no help.

Let us perform the following change of variables
\be X=\frac{1}{2}e^{2\alpha}\cosh(2\phi)~~,\hspace{1em}
Y=\frac{1}{2}e^{2\alpha}\sinh(2\phi)~, \ee
which represents a mapping of minisuperspace into the wedge $X>|Y|$.
The minisuperspace interval (corresponding to DeWitt's supermetric)
can be expressed in the new coordinates as
\be \mbox{d}\phi^{2}-\mbox{d}\alpha^2
=\frac{1}{X^2-Y^2}(\mbox{d}Y^{2}-\mbox{d}X^2)~, \ee with
$(X^2-Y^2)^{-1}$ the conformal factor\footnote{Recall that any two
  metrics on a two-dimensional manifold are related by a conformal
  transformation.}. As an immediate consequence of conformal
invariance, uniformly expanding (contracting) Universes are given by
straight lines parallel to $X=Y$ ($X=-Y$) within the wedge. The
``horizons'' $X=-Y$ and $X=Y$ represent the initial and final
singularity, respectively.  It is worth pointing out that the
classical trajectory Eq.~(\ref{Classical Solution}) takes now the much
simpler expression
\be \overline{k}=2X~, \ee 
\emph{i.e.} classical trajectories are
represented by straight lines parallel to the $Y$ axis and with the
extrema on the two singularities.

In the new coordinates $X, Y$ the operator $\hat{T}_{0}$ in
Eq.~(\ref{constraint operator}) reads
\be\label{Conformal to KG}
\hat{T}_{0}=(X^2-Y^2)(\partial_X^2-\partial_Y^2-4)~,
\ee
which is, up to the inverse of the conformal factor, a Klein-Gordon
operator with $m^2=4$. This is a first step towards a perturbative
solution of Eq.~(\ref{first order perturbation}), which we rewrite
below for convenience of the reader in the form
\be\label{equation for the perturbation with U}
\hat{T}_{0}\psi^{(1)}=U(\alpha,\phi)\psi^{(0)}~,
\ee
where
\be
U(\alpha,\phi)=-e^{3\alpha}g(\alpha,\phi)~.
\ee
Note that the above is the same equation as the one considered in the
previous section, but we are now allowing for the interaction
potential to depend also on $\phi$. Given Eq.~(\ref{Conformal to KG}),
we recast Eq.~(\ref{equation for the perturbation with U}) in the form
that will be used for the applications of the next section, namely
\be\label{eq-above}
(\partial_X^2-\partial_Y^2-4)\psi^{(1)}=\frac{U(X,Y)}{(X^2-Y^2)}\psi^{(0)}~.
\ee
The formal solution to Eq.~(\ref{eq-above}) is given by a convolution
of the r.h.s with the Green's function satisfying suitably chosen
boundary conditions
\be\label{Convolution solution}
\psi^{(1)}(X,Y)=\int\mbox{d}X^{\prime}\mbox{d}Y^{\prime}\;
G(X,Y;X^{\prime},Y^{\prime})\frac{U(X^{\prime},Y^{\prime})}{(X^{\prime
    2}-Y^{\prime2})}\psi^{(0)}(X^{\prime},Y^{\prime})~.  \ee
The Green's  function of the Klein-Gordon operator in free space is
well-known for any dimension $D$ (see \emph{e.g.}
Ref.~\cite{Zhang}). In $D=2$ it is formally given by\footnote{Notice
  that here $Y$ plays the r\^ole of time.}
\be\label{Green function} G(X,Y;X^{\prime},Y^{\prime})=\int
\frac{\mbox{d}^2 k}{(2\pi)^2}\frac{e^{-i
    \left(k_Y(Y-Y^{\prime})-k_X(X-X^{\prime})\right)}}
{\left(k_{Y}^{2}-k_{X}^{2}\right)-4}~,
\ee
and satisfies the equation
\be\label{Green equation}
(\partial_X^2-\partial_Y^2-4)G(X,Y;X^{\prime},Y^{\prime})
=\delta(X-X^{\prime})\delta(Y-Y^{\prime})~.
\ee
Evaluating (\ref{Green function}) explicitly using Feynman's
integration contour, which is a preferred choice in the context of a
third quantization, we get~\cite{disessa, Zhang}
\be
G(X,Y;X^{\prime},Y^{\prime})=-\frac{1}{4}\theta(s)H_{0}^{(2)}(2\sqrt{s})-\frac{i}{2\pi}\theta(-s)K_{0}(2\sqrt{-s})~,
\ee
where we introduced the notation $s=(Y-Y^{\prime})^2-(X-X^{\prime})^2$
for the interval.  However, the present situation is distinguished
from the free case, since there is a physical boundary represented by
the edges of the wedge. The boundary conditions must be therefore
appropriately discussed. A preferred choice is the one that leads to
the Feynman boundary conditions in the physical coordinates
$(\alpha,\phi)$. In the following, we will see the form that these
conditions take in the new coordinate system, finding the
transformation laws of the operators that annihilate progressive and
regressive waves.

We begin by noticing that the generator of dilations in the $(X,Y)$
plane acts as a tangential derivative along the edges
\be\label{dilation}
X\partial_{X}+Y\partial_{Y}=\partial_{t}~.
\ee
Moreover on the upper edge $X=Y$ (corresponding to the big crunch) we have
\be
\partial_{t}|_{X=Y}=\frac{1}{4}\left(\partial_{\alpha}+\partial_{\phi}\right)~,
\ee
while on the lower edge (corresponding to the initial singularity) we have
\be
\partial_{t}|_{X=-Y}=\frac{1}{4}\left(\partial_{\alpha}-\partial_{\phi}\right)~.
\ee
Therefore the boundary conditions
\be
\partial_{t}|_{X=Y}G=\partial_{t}|_{X=-Y}G=0
\ee
are equivalent to the statement that the Green's function is a
positive (negative) frequency solution of the wave equation at the
final (initial) singularity. This is in agreement with the Feynman
prescription and with the fact that the potential $e^{\alpha}$
vanishes at the singularity $\alpha\to-\infty$.

There is a striking analogy with the classical electrostatics problem
of determining the potential generated by a point charge inside a
wedge formed by two conducting plates (in fact it is well-known that
the electric field is normal to the surface of a conductor, so that
the tangential derivative of the potential vanishes). The similarity
goes beyond the boundary conditions and holds also at the
level of the dynamical equation. In fact, after performing a Wick rotation $Y\to -i\,Y$
Eq.~(\ref{Green equation}) becomes the Laplace equation with a
constant mass term, while the operator $\partial_t$ defined in
Eq.~(\ref{dilation}) keeps its form.

\begin{figure}
\centering
\begin{minipage}{0.45\textwidth}
\centering
 \includegraphics[width=\columnwidth]{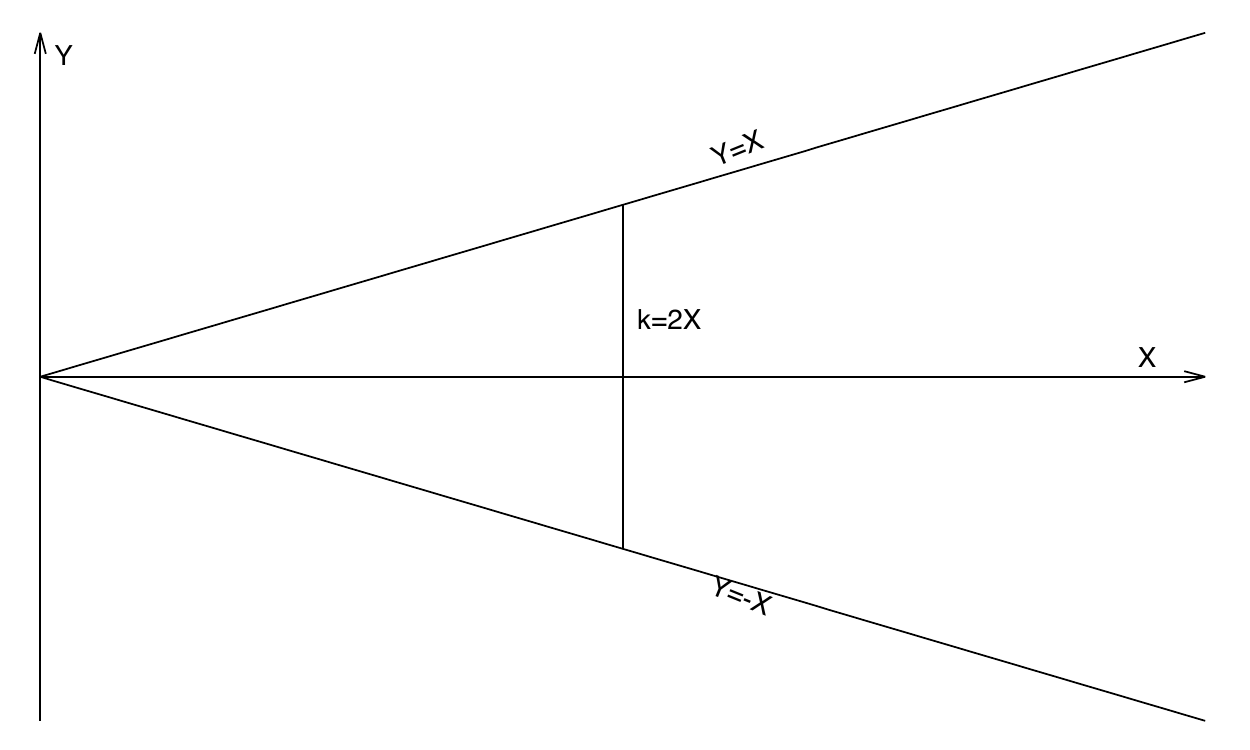}
\caption{The whole of minisuperspace is conformally mapped into the wedge. The straight lines $X=\pm Y$ correspond to the two horizons, while the vertical line is the classical trajectory of a closed Universe.}
\end{minipage}\hfill
\begin{minipage}{0.45\textwidth}
\centering
\hspace*{-1cm}\includegraphics[width=1.3\columnwidth]{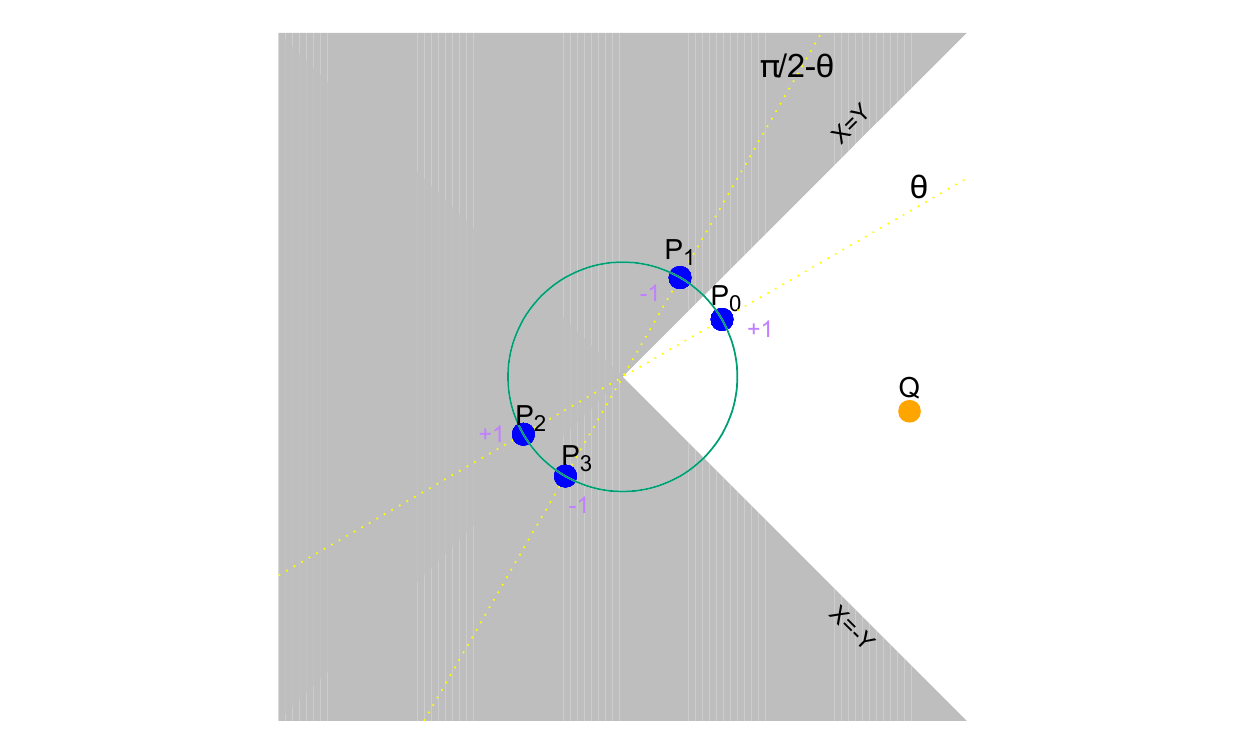}
\caption{The Green's function is a function of the point $Q$ and the source $P_0$. The positions and the charges of the images $P_j$'s are determined so as to satisfy the boundary conditions.}
\label{fig:wedge}
\end{minipage}
\end{figure}

After performing the Wick rotation the problem can be solved with the
method of images. Given a source (charge) at point
$P_0=(r^{\prime},\theta^{\prime})$, three image charges as in Fig. \ref{fig:wedge} are
needed to guarantee that the boundary conditions are met. The
Euclidean Green's function as a function of the point $Q=(r,\theta )$
and source $P_0$ (hence $m=2$) is shown to be given by
\be G(r,\theta ,r^{\prime},\theta^{\prime})=\frac{1}{2\pi}\left(K_0(m
\,\overline{P_0 Q})-K_0(m \,\overline{P_1 Q})+K_0(m \,\overline{P_2
  Q})-K_0(m \,\overline{P_3 Q})\right)~.  \ee
The quantities $\overline{P_j Q}$ represent the Euclidean distances
between the charges and the point $Q$. The Lorentzian Green's function
is then recovered by Wick rotating all the $Y$ time coordinates,
\emph{i.e.} those of $Q$ and of the $P_j$'s. In order to obtain this
solution we treated the two edges symmetrically, thus maintaining time
reversal symmetry. The Green's function with a source $P_0$ within the
wedge vanishes when $Q$ is on either of the two edges. In fact, this
can be seen as a more satisfactory
way of realizing DeWitt's boundary
condition, regarding it as a property of correlators rather than of states\footnote{DeWitt originally proposed the vanishing of the
  wave function of the Universe at singular metrics on superspace, suggesting in this way that the singularity problem would be solved \emph{a priori} with an appropriate choice of the boundary conditions.
 However, there are cases (see discussion in Ref.~\cite{Bojowald:2002, Bojowald:2003}) where DeWitt's proposal does not lead to a well posed boundary value problem and actually overconstrains the dynamics. For instance, the solution given in Ref.~\citep{Kiefer:1988} that we discussed in Section \ref{Section unperturbed} satisfies it for $\alpha\to\infty$ but not at the initial singularity, since all the elementary solutions $C_{k}(\alpha)$ are indefinitely oscillating in that region. Imposing the same
  condition on the Green's function does not seem to lead to such
  difficulties.}.
  
\section{Treating the interaction as white noise}\label{section noise}

The methods developed in the previous sections are completely general
and can be applied to any choice of the extra interaction terms using
Eq.~(\ref{Convolution solution}).  In this section we will consider,
as a particularly simple example, the case in which the additional
interaction is given by white noise. Besides the mathematical
simplicity there are also physical motivations for doing so.  In fact,
we can make the assumption that the additional terms that violate the
Hamiltonian constraint can be used to model the effect of the
underlying discreteness of spacetime on the evolution of the Universe.

Some approaches to Quantum Gravity (for instance GFT) suggest that an
appropriate description of the gravitational interaction at a
fundamental level is to be given in the language of third
quantization~\cite{Oriti:2013}. At this stage, one might make an
analogy with the derivation of the Lamb shift in Quantum
Electro-Dynamics (QED) and in effective stochastic approaches. As it is well known, the effect stems from
the second quantized nature of the electromagnetic field. However, the
same prediction can also be obtained if one holds the electromagnetic
field as classical, but impose \emph{ad hoc} conditions on the
statistical distribution of its modes, which necessarily has to be the same as
that corresponding to the vacuum state of QED. In this way, the
instability of excited energy levels in atoms and the Lamb shift are
seen as a result of the interaction of the electron with a
\emph{stochastic} background electromagnetic field~\cite{Lamb-shift-SED, Lamb-shift-spontaneous-emission-SED}. However, in the
present context we will not resort to a third quantization of the
gravitational field, but instead put in \emph{ad hoc} stochastic terms
arising from interactions with degrees of freedom other than the scale
factor. Certainly, our position here is weaker than that of Stochastic
Electro-Dynamics (SED) (see Ref.~\cite{de1983stochastic, QM-SED} and the works
cited above), since a fundamental third quantized theory of gravity
has not yet been developed to such an extent so as to make observable
predictions in Cosmology. Therefore we are unable to give details
about the statistical distribution of the gravitational degrees of
freedom in what would correspond to the vacuum state. Our model should
henceforth be considered as purely phenomenological and its link to
the full theory will be clarified only when the construction of the latter will eventually
be completed.

To be more specific, we treat the function $g$ in the perturbation as
white noise. Stochastic noise is used to describe the interaction of a
system with other degrees of freedom regarded as an
\emph{environment}\footnote{For a derivation of the
  Schr{\"o}dinger-Langevin equation using the methods of stochastic
  quantization we refer the reader to Ref.~\cite{Yasue1977}.  We are
  not aware of other existing works suggesting the application of
  stochastic methods to Quantum Cosmology. There are however
  applications to the fields of Classical Cosmology and inflation and
  the interested reader is referred to the published literature.}.
Hence we write
\begin{align}
\langle g(\alpha, \phi)\rangle&=0~,\\ \langle g(\alpha, \phi)
g(\alpha^{\prime}, \phi^{\prime}) \rangle &
=\delta(\alpha-\alpha^{\prime})\delta(\phi-
\phi^{\prime})~,\label{seconda proprieta' noise}
\end{align}
where $\langle \underline{\hspace{1em}}\rangle$ denotes an ensemble average.
It is straightforward to see that
\be
\langle\psi^{(1)}\rangle=0~,
\ee
which means that the ensemble average of the corrections to the wave
function vanishes.

A more interesting quantity is represented by the second moment
\be\label{definition variance}
\mathcal{F}(X_1,Y_1;X_2,Y_2)=
\langle (\psi^{(1)}(X_1,Y_1))^{*}\psi^{(1)}(X_2,Y_2)\rangle~.
\ee
In fact when evaluated at the same two points
$\tilde{\mathcal{F}}(X,Y)\equiv\mathcal{F}(X,Y;X,Y)$, it represents
the variance of the \emph{statistical} fluctuations of the wave
function at the point $(X,Y)$.  Using Eqs.~(\ref{first order
  perturbation}), (\ref{seconda proprieta' noise}) and
(\ref{definition variance}) we get \be\label{expression for the
  variance} \tilde{\mathcal{F}}(X,Y)=16\int_{X\geq |Y|}
\mbox{d}X^{\prime}\,\mbox{d}Y^{\prime}\;(X^{\prime 2}-Y^{\prime
  2})^{\frac{1}{2}}|\psi^{(0)}(X^{\prime},Y^{\prime})|^2 \,
|G(X-X^{\prime},Y-Y^{\prime})|^2~.  \ee
In order to obtain the correct expression of the integrand, one needs
to use the transformation properties of the Dirac distribution,
yielding
\be
\langle g(\alpha, \phi) g(\alpha^{\prime}, \phi^{\prime})\rangle
=2(X^2-Y^2)\delta(X-X^{\prime})\delta(Y-Y^{\prime})~.
\ee
Notice the resemblance of Eq.~(\ref{definition variance}) with the
two-point function evaluated to first order using Feynman rules for a
scalar field in two dimensions interacting with a potential
$(X^2-Y^2)^{\frac{1}{2}} |\psi^{(0)}|^2$. Following this analogy, the
variance $\tilde{\mathcal{F}}$ can be seen as a vacuum bubble.

Equation~(\ref{expression for the variance}) implies that a white noise
interaction is such that the different contributions to the modulus
square of the perturbations add up incoherently.


\section{Conserved charge and the choice of the inner product}\label{Section on the charge}
Let us briefly discuss the r\^ole of the Noether charge and the choice
of the inner product.  Equation (\ref{WDW equation}) is just a
Klein-Gordon equation (with $\phi$ interpreted as time\footnote{We
  point out that another reason to prefer $\phi$ as a time variable
  instead of $\alpha$ is that this choice leads to a potential that is
  bounded from below. Even from a classical perspective $\alpha$ is
  not a good choice as a \emph{global} time since it is not a
  monotonic function of the proper time. However, it can still be used
  as a local time (a consistent way of using local times was studied
  in Ref.~(\cite{Bojowald:2010xp}). In fact this is a problem one has
  to deal with for general, non-deparameterizable systems.}) with a
potential $V(\alpha)=e^{4\alpha}-k^2$.

Equation (\ref{WDW equation}) can be seen as the Euler-Lagrange equation of the Lagrangian
\be
\mathcal{L}={\partial\psi^{*}\over\partial\alpha}{\partial\psi\over\partial\alpha}-{\partial\psi^{*}\over\partial\phi}{\partial\psi\over\partial\phi}+e^{4\alpha}\psi^{*}\psi~.
\ee
Since the Lagrangian is manifestly $U(1)$ invariant, the following
conservation law holds as a consequence of Noether's theorem
\begin{align}
\partial_{\mu}j^{\mu}=0,
\end{align}
where\footnote{We introduced a compact notation to denote the
  antisymmetric combination of right and left derivatives, namely
$$
   {\overset{\leftrightarrow}{\partial}\over\partial\phi}
\equiv{\overset{\rightarrow}{\partial}\over\partial\phi}-
{\overset{\leftarrow}{\partial}\over\partial\phi}
   $$
   }
\be\label{Noether current}
j_{\mu}=-i\left(\psi^{*}{\overset{\leftrightarrow}{\partial}
\psi\over\partial\phi},\;\psi^{*}{\overset{\leftrightarrow}
{\partial}\psi\over\partial\alpha}\right).
\ee
The corresponding Noether charge (obtained following the standard
procedure) reads
\be\label{Noether charge}
Q=\int_{-\infty}^{\infty}\mbox{d}\alpha\; j_{\phi}~, \ee
and is
conserved under time evolution 
\be\label{Charge conservation}
{\mbox{d}Q\over\mbox{d}\phi}=0~.  \ee
Clearly, the conservation of $Q$ is in general incompatible with
 that of the $L^2$ norm of $\psi$, defined by
\be
\lVert\psi\rVert_{L^2(\mathbb{R})}=\int_{-\infty}^{\infty}\mbox{d}\alpha\;
|\psi|^2~.  \ee
Furthermore, the Noether charge $Q$ has no definite sign unless one
considers only solutions with either positive or negative frequency,
since a generic state admits a decomposition in terms of positive and
negative frequency solutions of the free WDW equation. Such solutions
span superselection sectors that are preserved by the Dirac
observables identified in Ref.~\cite{Ashtekar:2006D73}, and states
in a given superselection sector satisfy a Schr{\"o}dinger equation
with a Hamiltonian $\pm\sqrt{\Theta}$. Within each sector, the
conservation of the Hilbert norm holds true and Eq.~(\ref{Charge
  conservation}) is an obvious consequence of energy conservation. As
far as the free dynamics is concerned, it is perfectly legitimate to
work in a superselection sector and use it to construct the
(kinematical) Hilbert space\footnote{The relation between different
  choices for the physical inner product and the Green's functions in
  LQC was explored in Ref.~\cite{Calcagni:2010ad}.}~\citep{Ashtekar:2006, Ashtekar:2006D73}. However, when the
theory includes either non-linear interactions or potentials that
depend on (internal) time, superselection no longer holds and both
sectors have to be taken into account, thus naturally leading to a
third quantization~\cite{Isham:1992}.  The non-equivalence of $Q$ and
$\lVert\psi\rVert_{L^2(\mathbb{R})}$ can be easily seen for a wave
packet (\ref{Wave Packet}), in which case one has 
\be
Q=\int_{-\infty}^{\infty}\mbox{d}k\; 2k|A(k)|^2~.  \ee 
Note that when $A(k)$ is supported only on one side of the real line,
Eq.~(\ref{Charge conservation}) can be interpreted as a conservation
law for the average momentum\footnote{In the QFT context the same
  relation is interpreted as a relativistic normalization, which
  amounts to having a density of $2k$ particles per unit volume.}. A
similar interpretation is not available instead when both positive
frequency and negative frequency solutions are present in the
decomposition of the wave packet.  For the reasons given above, a
naive probabilistic interpretation of the theory along the lines of
ordinary Quantum Mechanics does not seem possible in general. It is
thus preferable to avoid using the designation of wave function for
the classical field $\psi$.

The Noether charge of the semiclassical Universe considered in
Section~\ref{Section unperturbed} can actually be computed
analytically, and since it is conserved one can evaluate it in the
simplest case, \emph{i.e.} when the potential vanishes for $\alpha\to
-\infty$. In fact when both $\alpha$ and $\phi$ are in a neighborhood
of $-\infty$, \emph{i.e.} close to the initial singularity, one can
approximate the exact solution with a Gaussian wave packet (given by
the WKB approximation, see Appendix \ref{WKB solution}), which
describes the evolution of the Universe in the expanding phase. We obtain
\be
\psi\overset{\mathsmaller{\alpha\to-\infty}}{\mathlarger{\approx}}
c_{\overline{k}}\,b\sqrt{\frac{\pi}{2}}\exp\left\{\frac{i}{2}\left[\overline{k}\left(2\phi +\arccosh\left(\frac{\overline{k}}{e^{2\alpha}}\right)+\sqrt{1-\frac{e^{4\alpha}}{\overline{k}^2}}\right)-\frac{\pi}{2}
\right]\right\}
\exp\left[-\frac{b^2}{2}\left(\phi+\frac{1}{2}\arccosh\left({\overline{k}\over
    e^{2\alpha}}\right)\right)^2\right], \ee 
and then using Eq.~(\ref{Noether
  current}) the charge density reads 
\be
j_{\phi}=|c_{\overline{k}}|^2 \pi b^2 \overline{k}\exp\left[-{b^2\over
    4}\left(2\phi+\arccosh\left({\overline{k}\over
    e^{2\alpha}}\right)\right)^2\right]~.  \ee 
For large negative values of $\alpha$ one can approximate the inverse
hyperbolic function in the argument of the exponential as
\be \arccosh\left({\overline{k}\over
  e^{2\alpha}}\right)\overset{\mathsmaller{\alpha\to-\infty}}
{\mathlarger{\approx}}\log(2\overline{k})-2\alpha~,
\ee 
and hence reduce the evaluation of the charge to that of a Gaussian
integral, leading to the result
\be
Q=|c_{\overline{k}}|^2 b\,\pi^{3/2}\; \overline{k}~.  \ee 

From the above analysis we conclude that the quantum conservation law
leads, in the classical limit, to momentum conservation. We recall
that classically the momentum is related to the derivative of the
scalar field w.r.t. proper time, as
\be
\overline{k}=a^3\dot{\phi}~.
\ee
%

\section{Conclusions}

Motivated by LQC and GFT we considered an extension of WDW
minisuperspace cosmology with additional interaction terms representing a self-interaction of the Universe. Such terms can be seen as a particular case of the model considered in Ref.~\cite{Calcagni:2012}, which was proposed as an approach to quantum dynamics of inhomogeneous cosmology. In general, inhomogeneities would lead to non-linear differential equations for the quantum field. This is indeed the case when the ``wave function'' of the Universe is interpreted as a quantum field or even as a classical field describing the hydrodynamics limit of GFT.

In the framework of a first quantized cosmology, we only considered linear modifications of the theory in order to secure the validity of the superposition principle. Our work represents a first step towards a more general study, that should take into account non-linear and possibly non-local interactions in minisuperspace. However, the way such terms arise, their exact form, and even the precise way in which minisuperspace dynamics is derived from fundamental theories of Quantum Gravity should be dictated from the full theory itself.

Assuming that the additional interactions are such that deviations from the Friedmann equation are small, we developed general perturbative methods which allowed us to solve the modified WDW equation. We considered a closed FLRW Universe filled with a massive scalar field to define an internal time, for which wave packets solutions are known explicitly and propagate with no dispersion.
A modified WDW equation is then obtained in the large volume limit of a particular GFT inspired extension of LQC for a closed FLRW Universe. Perturbative methods are then used to find the corrections given by self-interactions of the Universe to the exact solution given in Ref.~\cite{Kiefer:1988}. To this end, the Feynman propagator of the WDW equation is evaluated exactly by means of a conformal map in minisuperspace and (after a Wick rotation) using the method of the image charges that is familiar from electrostatics. This is potentially interesting as a basic building block for any future perturbative analysis of non-linear minisuperspace dynamics for a closed Universe.

A Helmoltz-like equation was obtained from WDW when the extra interaction does not depend on the internal time. Its Green's kernel was evaluated exactly and turned out to depend on a free parameter $\eta$ related to the choice of boundary conditions. Further research must be carried over to link $\eta$ to different boundary proposals.

We illustrated our perturbative approach in the
simple and physically motivated case in which the
perturbation is presented by white noise. In this phenomenological
model the stochastic interaction term can be seen as describing the
interaction of the cosmological background with other degrees of freedom
of the gravitational field. Calculating the variance of the
statistical fluctuations of the wave function, we found that a white
noise interaction is such that the different contributions to the
modulus square of the perturbations add up incoherently.

\begin{appendices}
\section{WKB approximation of the elementary solutions and asymptotics of the wave packets}\label{WKB solution}

Since Eq.~(\ref{EQUATION FOR ALPHA}) has the form of a time-independent
Sch\"rodinger equation, it is possible to construct approximate
solutions using the WKB method.

For a given $k$, we divide the real line in three regions, with a
neighborhood of the classical turning point in the middle. The
classical turning point is defined as the point where the potential is
equal to the energy
\be
\alpha_{k}=\frac{1}{2}\log{k}~.
\ee
The WKB solution to first order in the classically allowed region
$]-\infty, \alpha_{k}-\epsilon]$ (with $\epsilon$ an appropriately
    chosen real number, see below) is 
\be\label{allowed region}
C_{k}^{I}=\frac{2}{(k^2-e^{4\alpha})^{1/4}}\cos\left(\frac{1}{2}\sqrt{k^2-e^{4\alpha}}-\frac{k}{2}\mbox{arccoth}\left(\frac{k}{\sqrt{k^2-e^{4\alpha}}}\right)+\frac{\pi}{4}\right)~,
\ee
and reduces to a plane wave in the allowed region for
$\alpha\ll\alpha_{k}$. The presence of the barrier fixes the amplitude
and the phase relation of the incoming and the reflected wave through
the matching conditions.

The solution in the classically forbidden region $[
  \alpha_{k}+\epsilon,+\infty[ $ is instead exponentially decreasing
    as it penetrates the potential barrier and reads
\be\label{forbidden region}
C_{k}^{III}=\frac{1}{(e^{4\alpha}-k^2)^{1/4}}\exp\left(-\frac{1}{2}\sqrt{e^{4\alpha}-k^2}+\frac{k}{2}\arctan\left(\frac{k}{\sqrt{e^{4\alpha}-k^2}}\right)\right)~.
\ee
Finally, in the intermediate region
$]\alpha_{k}-\epsilon,\alpha_{k}+\epsilon[$ any semiclassical method
    would break down, hence the Schr\"odinger equation must be solved
    exactly using the linearized potential
\be
V(\alpha)\simeq V(\alpha_{k})+V^{\prime}(\alpha_{k})(\alpha-\alpha_{k})=4e^{4\alpha_{k}}(\alpha-\alpha_{k})=4k^2(\alpha-\alpha_{k})~.
\ee
In this intermediate regime, Eq.~(\ref{EQUATION FOR ALPHA}) can be
rewritten as the well-known Airy equation
\be\label{Airy equation}
\frac{\mbox{d}^2 C_{k}^{II} }{\mbox{d}t^2}-t C_{k}^{II}=0~,
\ee
where the variable $t$ is defined as
\be
t=(4k^2)^{1\over 3}(\alpha-\alpha_{k})~.
\ee
Of the two independent solutions of Eq.~(\ref{Airy equation}),
only the Airy function $\mbox{Ai}(t)$ satisfies the boundary
condition, and hence we conclude that
\be
C_{k}^{II}=c \; Ai(t)~,
\ee
where $c$ is a constant that has to be determined by matching the
asymptotics of $\mbox{Ai}(t)$ with the WKB approximations on both
sides of the turning point. Hence we get that, for $t\ll0$
\be
\mbox{Ai}(t)\approx\frac{\cos\left(\frac{2}{3}|t|^{3/2}-\frac{\pi}{4}\right)}{\sqrt{\pi}|t|^{1/4}}~,
\ee while for $t\gg0$ 
\be
\mbox{Ai}(t)\approx\frac{e^{-\frac{2}{3}t^{3/2}}}{2\sqrt{\pi}t^{1/4}}~.
\ee 
We thus fix $c=2\sqrt{\pi}$.
Note that the arbitrariness in the choice of $\epsilon$ can be solved,
\emph{e.g.} by requiring the point $\alpha_{k}-\epsilon$ to coincide
with the first zero of the Airy function.

  \begin{figure}\label{tre curve WKB}
  \centering
 \includegraphics[width=\textwidth]{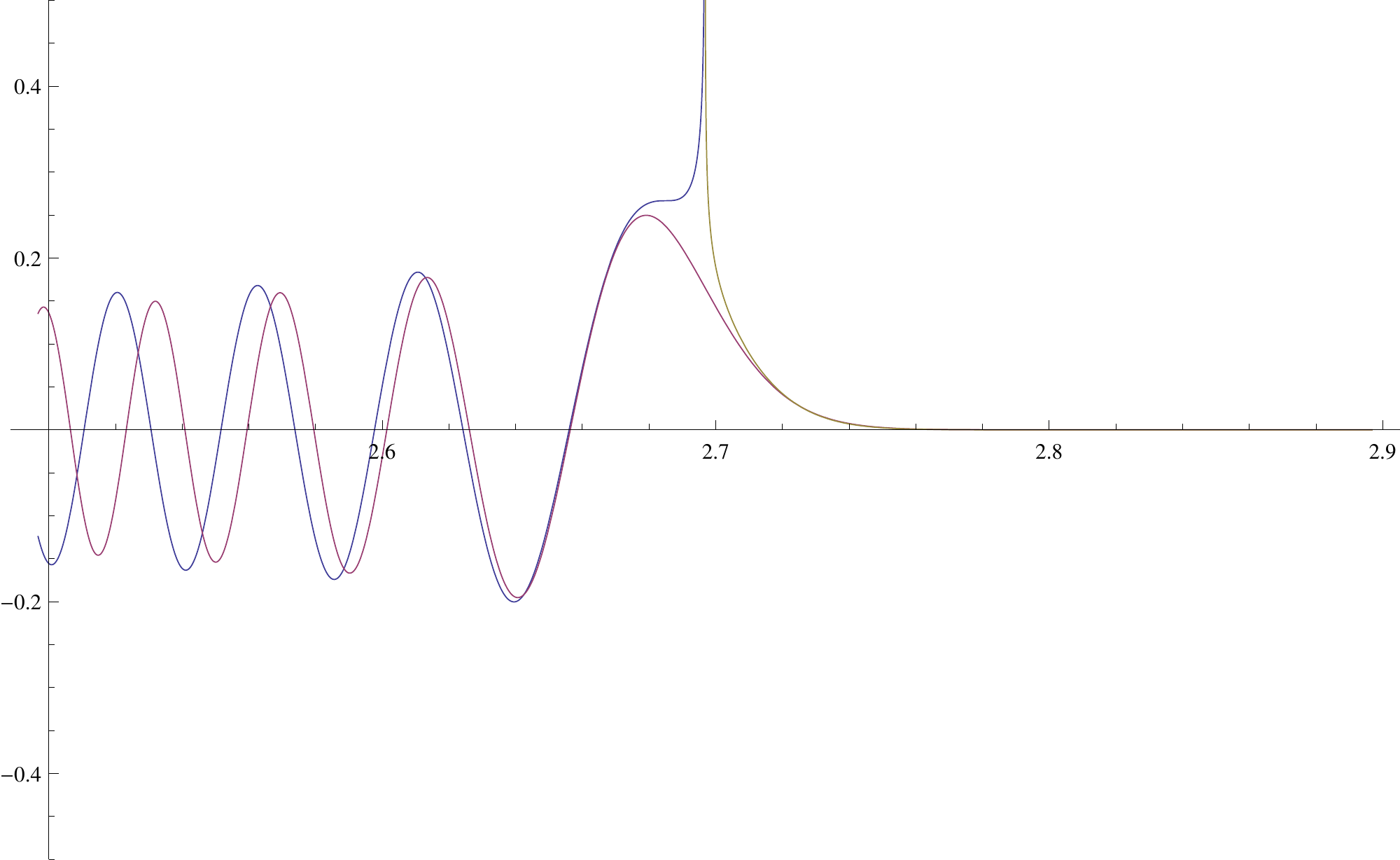}
\caption{The blue and the yellow curve represent, the
  WKB approximation of the wave function in the classically allowed
  and in the forbidden region, respectively. They both diverge at the classical
  turning point. The Airy function is represented in red. The value
  $k=220$ was chosen.}
    \end{figure}
The WKB approximation improves at large values of $k$, as one should
expect from a method that is semiclassical in spirit. Yet it allows
one to capture some effects that are genuinely quantum, such as the
barrier penetration and the tunneling effect.

From the approximate solution we have just found, we can construct
wave packets as in Ref.~\cite{Kiefer:1988}.
We thence restrict our attention to the classically allowed region and
the corresponding approximate solutions, \emph{i.e.}
$C_{k}^{I}(\alpha)$ for $\alpha\ll\alpha_{k}=\frac{1}{2}\log{k}$ and
compute the integral in Eq.~(\ref{Wave Packet}). If the Gaussian
representing the amplitudes of the monochromatic modes is narrow
peaked, \emph{i.e.} its variance $b^2$ is small enough, we can
approximate the amplitude in Eq.~(\ref{allowed region}) with that
corresponding to the mode with the mean frequency
$\overline{k}$. Thus, introducing the constant
\be
c_{\overline{k}}=\frac{2}{(\overline{k}^2-e^{4\alpha})^{1/4}}\frac{1}{\pi^{1/4}\sqrt{b}}~,
\ee
we have
\be\label{stima}
\psi(\alpha,\phi)\simeq c_{\overline{k}}\int_{-\infty}^{\infty}\mbox{d}k\; e^{-\frac{(k-\overline{k})^2}{2b^2}}\cos\left(\frac{1}{2}\sqrt{k^2-e^{4\alpha}}-\frac{k}{2}\mbox{arccoth}\left(\frac{k}{\sqrt{k^2-e^{4\alpha}}}\right)+\frac{\pi}{4}\right)e^{ik\phi}~.
\ee
Moreover,
\be\label{dominant term}
k\,\mbox{arccoth}\left(\frac{k}{\sqrt{k^2-e^{4\alpha}}}\right)=k\arccosh\frac{k}{e^{2\alpha}}\simeq k\arccosh\frac{\overline{k}}{e^{2\alpha}}~.
\ee
The last approximation in the equation above holds as the derivative of
the inverse hyperbolic function turns out to be much smaller than unity
in the allowed region. Furthermore, the term approximated in
Eq.~(\ref{dominant term}) dominates over the square root in the
argument of the cosine in the integrand in the r.h.s. of
Eq.~(\ref{stima}), so we can consider the latter as a constant. Hence,
we can write
\be\label{psi}
\psi(\alpha,\phi)\simeq c_{\overline{k}}\int_{-\infty}^{\infty}\mbox{d}k\; e^{-\frac{(k-\overline{k})^2}{2b^2}}\cos\left(\Lambda k+\delta\right)e^{ik\phi}~,
\ee
where we have introduced the notation
\begin{align}
\Lambda&\equiv\frac{1}{2}\arccosh\frac{\overline{k}}{e^{2\alpha}}~,\\
\delta&\equiv\frac{1}{2}\sqrt{\overline{k}^2-e^{4\alpha}}-\frac{\pi}{4}~.
\end{align}
for convenience.
Using Euler's formula we can express the cosine in the integrand in the r.h.s. of 
Eq.~(\ref{psi}) in terms of complex exponentials and evaluate $\psi(\alpha,\phi)$.
In fact, defining
\be
P_{\overline{k}}\equiv e^{i\left(\overline{k}(\Lambda+\phi)+\delta\right)}~,~\hspace{1em}Q_{\overline{k}}\equiv e^{-i\left(\overline{k}(\Lambda-\phi)+\delta\right)}~,
\ee
we have
\be
\psi(\alpha,\phi)\simeq \frac{c_{\overline{k}}}{2}\int_{-\infty}^{\infty}\mbox{d}k\; e^{-\frac{(k-\overline{k})^2}{2b^2}}\left(e^{i(k-\overline{k})(\Lambda+\phi)}P_{\overline{k}}+e^{-i(k-\overline{k})(\Lambda-\phi)}Q_{\overline{k}}\right)~.
\ee
Shifting variables and performing the Gaussian integrations we get the
result as in Ref.~\cite{Kiefer:1988} 
\be \psi(\alpha,\phi)\simeq
c_{\overline{k}}\sqrt{\frac{\pi}{2}}b\left(e^{-\frac{b^2}{2}(\Lambda+\phi)^2}P_{\overline{k}}+e^{-\frac{b^2}{2}(\Lambda-\phi)^2}Q_{\overline{k}}\right)~.
\ee

\section{Orthogonality of the MacDonald functions}
It was proved in Refs.~\cite{Szmytkowski, Yakubovich, Passian} that
\be
\int_{0}^{\infty}\frac{\mbox{d}x}{x}\;K_{i\nu}(kx)K_{i\nu^{\prime}}(kx)=\frac{\pi^2\delta(\nu-\nu^{\prime})}{2\nu\sinh(\pi\nu)}~,
\ee
which expresses the orthogonality of the MacDonald functions of
imaginary order. Performing a change of variables, the above formula
can be recast in the form
 \be
\int_{-\infty}^{\infty}\mbox{d}\alpha\;
K_{i\frac{h}{2}}\left(\frac{e^{2\alpha}}{2}\right)K_{i\frac{h^{\prime}}{2}}\left(\frac{e^{2\alpha}}{2}\right)=\frac{\pi^2\delta(h-h^{\prime})}{h\sinh(\pi\frac{h}{2})}~,
\ee
which is convenient for the applications considered in this work.
For large values of $k$ it is equivalent to the normalization used
in Section~\ref{section 2}
\be \int_{-\infty}^{\infty}\mbox{d}\alpha
\;\psi^{*}_{k^{\prime}}(\alpha,\phi)\psi_{k}(\alpha,\phi)=\frac{\pi^2}{2}\delta(k-k^{\prime})~.
\ee
%


\end{appendices}

\bibliographystyle{h-physrev}
\bibliography{biblio1}

\end{document}